\documentclass{elsarticle}

\usepackage{graphicx}
\usepackage{color}
\usepackage{natbib}
\usepackage{amssymb}
\usepackage{bm}
\usepackage{amsfonts}
\usepackage{txfonts}
\usepackage{bbm}
\usepackage{bbold}

\def\spose#1{\hbox to 0pt{#1\hss}}

\def\lta{\mathrel{\spose{\lower 3pt\hbox{$\mathchar"218$}}
     \raise 2.0pt\hbox{$\mathchar"13C$}}}
\def\gta{\mathrel{\spose{\lower 3pt\hbox{$\mathchar"218$}}
     \raise 2.0pt\hbox{$\mathchar"13E$}}}
\newcommand{\ex}{\mathrm{e}}
\newcommand{\dd}{\mathrm{d}}

\newcommand{\ie}{\textsl{i.e.~}}
\newcommand{\cf}{\textsl{cf.~}}
\newcommand{\eg}{\textsl{e.g.~}}
\newcommand{\etal}{\textsl{et al.~}}

\newcommand{\etc}{\textsl{etc.~}}

 \newcommand{\cs}{{\Upsilon}}

\def\beq{\begin{equation}}
\def\eeq{\end{equation}}
\def\bea{\begin{eqnarray}}
\def\eea{\end{eqnarray}}
\def\eqref{\ref}

\def\n{\mathrm{n}}
\def\s{\mathrm{s}}

\def\LX{{\cal X}^\mu}
\def\LY{{\cal Y}^\mu}

\def\cn{c_\n}
\def\cs{c_\s}

\def\Cns{\mathcal{C}_{\n \s}}
\def\Csn{\mathcal{C}_{\s \n}}

\def\setR{\mathbb{R}}
\def\setC{\mathbb{C}}

\begin{document}

\begin{frontmatter}

\title{Cosmological Two-stream Instability}

\author{G.~L.~Comer}
\address{Department of Physics \& Center for Fluids at All Scales, Saint 
Louis University, St.~Louis, MO, 63156-0907, USA}

\author{P.~Peter}
\address{${\cal G}\setR\varepsilon\setC{\cal O}$ -- Institut 
d'Astrophysique de Paris, UMR7095 CNRS, Universit\'e Pierre \& Marie Curie, 
98 bis boulevard Arago, 75014 Paris, France}

\author{N.~Andersson}
\address{School of Mathematics, University of Southampton,
Southampton SO17 1BJ, UK}

\date{\today}

\begin{abstract}
Two-stream instability requires, essentially, two things to operate: a relative flow 
between two fluids and some type of interaction between them.  In this letter we 
provide the first demonstration that this mechanism may be active in a 
cosmological context. Building on a recently developed formalism for 
cosmological models with two, interpenetrating fluids with a relative flow between 
them, we show that two-stream instability may be triggered during the transition 
from one fluid domination to the other. We also demonstrate that the cosmological 
expansion eventually shuts down the instability by driving to zero the relative
flow and the coupling between the two fluids.  
\end{abstract}


\end{frontmatter}


Two-stream instability of two, interpenetrating plasmas is a well-established
phenomenon \cite{farley63:_2stream,buneman63:_2stream}. It is expected also 
to play a role in two, interpenetrating fluids, such as superfluid helium 
\cite{andersson04:_twostream}, gravitational-wave instabilities in neutron stars 
\cite{na03:_rev}, and glitches in pulsars (due to the interior superfluid 
components) \cite{acp03:_twostream_prl,andersson04:_twostream}. It is not 
surprising that the instability shows up in diverse settings, since the basic 
requirements for it to operate are fairly generic: there must be a relative flow and 
some type of coupling between the two fluids. A window of instability may be 
opened when a perturbation of the fluids (hereafter, ``mode'') appears to be 
``left-moving'', say, with respect to one fluid, but ``right-moving'' with respect to the 
other. The mechanism may be well-known, yet it was considered in a relativistic 
context only recently. Samuelsson \etal \citep{SLAC10} demonstrated quite 
generally the existence of (local) two-stream instability for a system of two, 
general relativistic fluids (using only causality -- mode-speed less than one, in 
geometric units -- and absolute stability -- modes are real for the fluids at rest -- to 
constrain the fluid properties).  In this letter we consider a new application by 
demonstrating that two-stream instability may be triggered in cosmological 
settings.

Although we use a specific example in this letter to exhibit the possibility of the 
mechanism to operate, we expect it to apply in various situations such as the 
presence of many inflaton scalar fields: the scalar fields turn into Bose-Einstein 
condensates behaving, for example, like the superfluids in a glitching neutron 
star, thus leading to two flows that have no particular reason to be aligned. 
Similarly, around the transition between dark matter and dark energy domination, 
the instability would provide a means of discriminating between a cosmological 
constant and a dark energy fluid.

Observations over the past few decades have provided a wealth of information 
that can be used to constrain cosmological models \citep{PPJPU}. One of the 
more stringent is the leading-order observed homogeneity and isotropy of the 
Universe. Is it possible to have a relative flow in cosmologically important settings 
that would not have been detected yet? An ideal stage is during a cosmological 
transition between one phase of domination to another. The requirement of a 
transition stems from the fact that one should naturally arrive at the current state of 
the Universe. We have shown, in a companion paper \citep{cpaprd}, that when 
one fluid flux dominates over the other (\ie before and after the transition), one 
recovers the usual Friedmann-Lema\^{\i}tre-Robertson-Walker (FLRW) behavior. 
However, during the transition the model goes through a Bianchi I phase.  We are 
thus led to ask whether the two-stream instability can be triggered during the 
same epoch? As we will show, the answer is ``yes". As a proof-of-principle we 
consider a fairly simplified picture, leaving actual cosmological consequences 
and constraints on more elaborate models (that deserve further examination) for 
future work.

\begin{figure*}[t]
\centering
\includegraphics[height=7.0cm,width=8.0cm,clip]{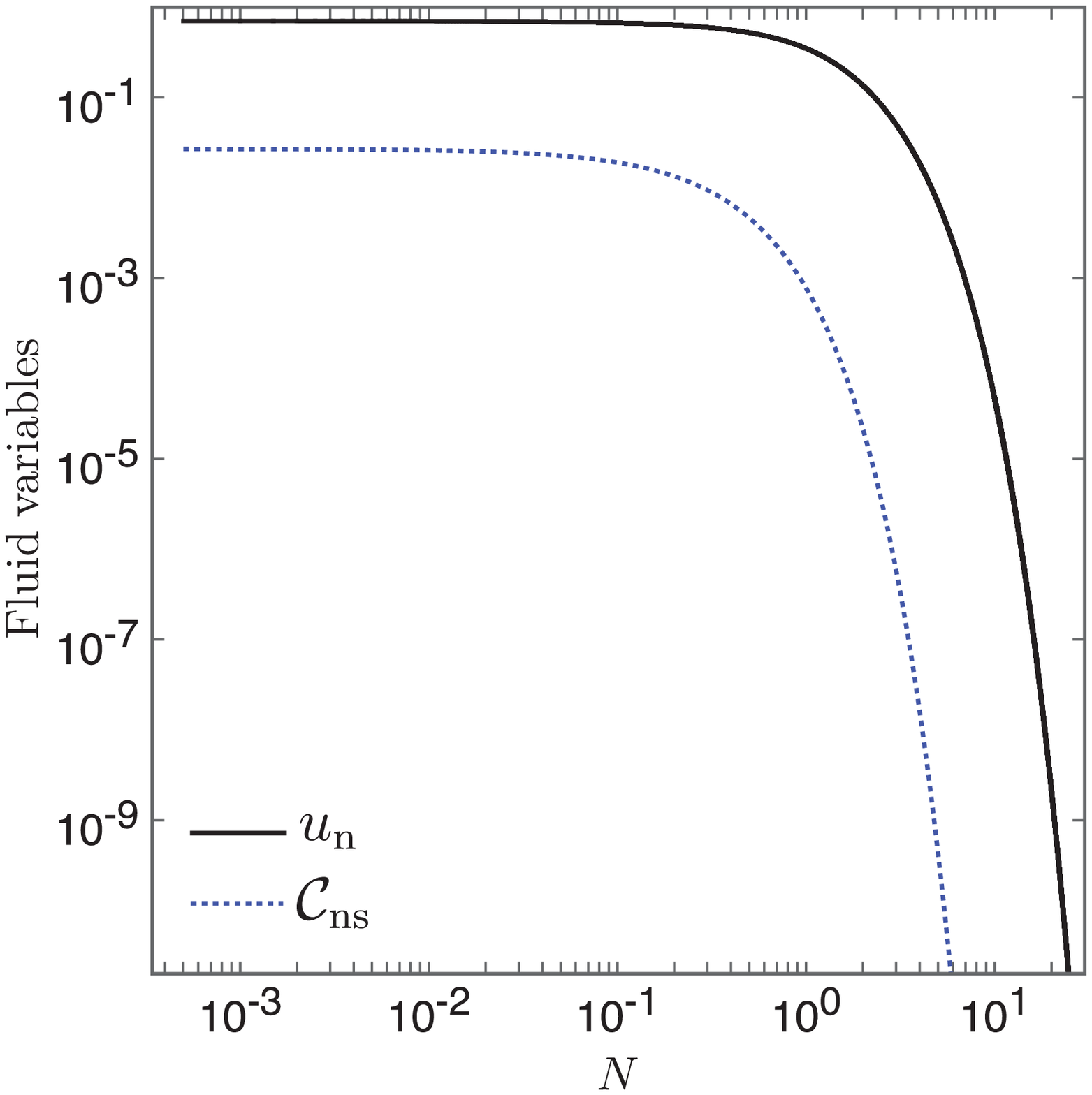} \hskip7mm
\includegraphics[height=7.0cm,width=8.0cm,clip]{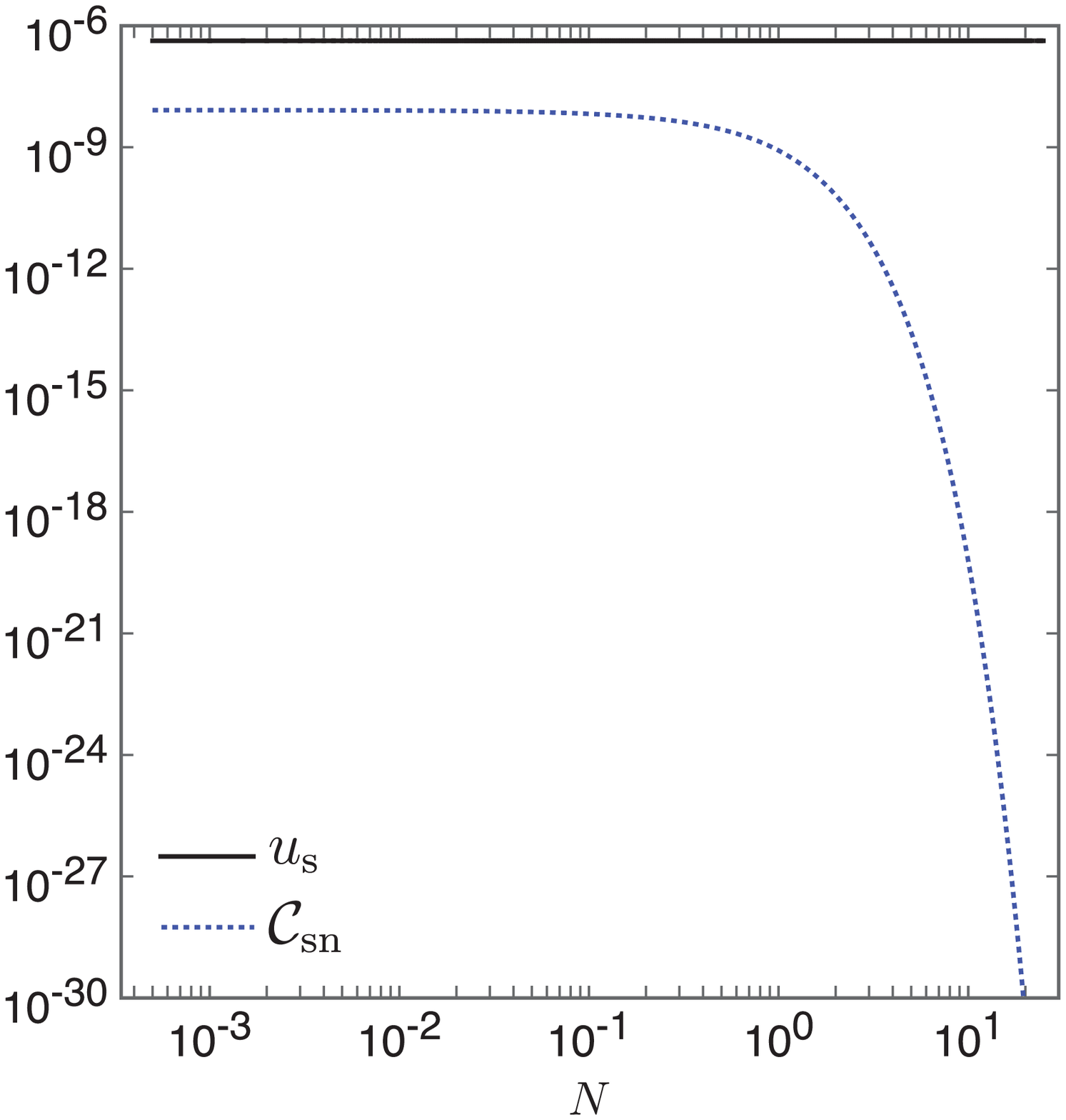}
\caption{Time evolution of the cosmological background quantities $u_\n$ and 
$\Cns$ (upper), and $u_\s$ and $\Csn$ (lower), as functions of the e-fold number
$N$ for $\alpha=1$ and $\beta=4/3$ (matter to radiation transition). With these 
values of $\alpha$ and $\beta$, the parameter $\kappa_\s$ is dimensionless 
while $\tau_{\n \s}$ has the dimensions of $m^{4-3(\sigma_\n - \sigma_\s)}$ (all 
functions and variables can be made dimensionless by means of a proper 
rescaling with some power of $m$). For all the graphs, we use $\kappa_\s = 1$, 
$\tau_{\n \s} = 0.1 m^{4-3(\sigma_\n - \sigma_\s)}$, $\sigma_\n = 1.1$, and 
$\sigma_\s = 1.1$. The initial values for the evolutions are 
$n(0) = 3.9\times 10^{-7}$, $s(0) = 1$, $u_\n (0)= 0.99$, 
$u_\s (0) = -4.25\times 10^{-7}$, $A_i(0) =2$, and $H_i(0) = 2.89$ (for each 
$i \in \{x,y,z\}$).}
\label{fig:uC}
\end{figure*}

Our cosmological two-fluid model has its relative flow along one direction, which 
we take to be the $z$-axis, \cf  \citep{cpaprd}.  Orthogonal to the flow we impose 
two, mutually orthogonal spacelike Killing vector fields: one along the $x$-axis 
and another along the $y$-axis.  The Killing vector fields $\LX$ and $\LY$ are 
thus $\LX = (0,1,0,0)$ and $\LY = (0,0,1,0)$. These two symmetries, and the 
remaining freedom in the choice of coordinates, imply the metric and fluid 
variables are functions of $z$ and time $t$. The metric can therefore be written  
\begin{eqnarray}
    \dd s^2 = - \dd t^2 + A_x^2 \dd x^2 + A_y^2 \dd y^2 + A_z^2 \dd z^2,
\label{fin_metric}
\end{eqnarray}
where the (still arbitrary) dimensionless functions $A_i$ ($i \in \{x,y,z\}$) can in 
principle depend on both $t$ and $z$.

To demonstrate two-stream instability, it is sufficient to consider fixed ``moments'' 
of cosmological time (\eg assume that the time-scale for the oscillations is much 
less than that of the cosmological expansion). For our mode analysis, this means 
any time derivatives of the background metric and the corresponding fluid flux will 
be ignored.  We will also ignore the metric perturbations and relax the 
$z$-dependence in the background metric and fluid flow so that our modes 
propagate in a spacetime of the well-known Bianchi I type. General cosmological 
perturbations in such a spacetime can be found in \citep{Zlosnik11}.

As described in \citep{cpaprd}, we use the multi-fluid formalism originally due to 
Carter \citep{carter89:_covar_theor_conduc} (see \citep{andersson07:_livrev} for 
a review). The fluid variables are two conserved four-currents, to be denoted 
$\displaystyle n^\mu = n u^\mu_\n$ (with 
$\displaystyle g_{\mu\nu} u^\mu_\n u^\nu_\n= - 1$) and
$\displaystyle s^\mu = s u^\mu_\s$ (with also 
$\displaystyle g_{\mu\nu} u^\mu_\s u^\nu_\s= - 1$). The fluid equations of motion 
are obtained from a Lagrangian $\Lambda(n,s)$, which we take to be of the form 
of a fluid with constituent mass $m$ coupled to a fluid with zero constituent mass; 
\ie
\begin{equation}
    \Lambda(n,s) = - m n^\alpha - \tau_{\n \s} n^{\sigma_\n} s^{\sigma_\s} - 
                   \kappa_\s s^\beta \ , \label{lam_mod}
\end{equation}
where $m$, $\alpha$, $\sigma_\n$, $\sigma_\s$, $\tau_{\n \s}$, $\kappa_\s$ and 
$\beta$ are constants.  The ``bare'' sound speeds \citep{SLAC10} are given by 
\begin{equation}
    \cn^2 \equiv \frac{\partial \ln \mu}{\partial \ln n}, \quad\quad
    \cs^2 \equiv \frac{\partial \ln T}{\partial \ln s},
\label{baress}
\end{equation}
where $\mu\equiv -\partial\Lambda/\partial n$ and 
$T\equiv -\partial\Lambda/\partial s$ are the associated conjugate 
momenta\footnote{We use notation that is reminiscent of the matter and entropy 
flows where $n$ is the total matter number density, $s$ the entropy density, 
$\mu$ the chemical potential, and $T$ the temperature.} of the two fluids. The 
cross-constituent coupling reads
\begin{equation}
    \Cns \equiv \frac{\partial \ln \mu}{\partial \ln s} 
              = \frac{T s}{\mu n} \Csn .
\label{cccoup}
\end{equation}

Performing perturbations to linear order, the fluid densities and the 
$z$-component of the unit four-velocities take the form
\begin{eqnarray}
    \bar{u}^z_{\n,\s}(t,z) &=& u^z_{\n,\s}(t) + \delta U^z_{\n,\s} 
                               \ex^{i k_\mu x^\mu} \ , \cr
    \bar{n}(t,z) &=& n(t) + \delta {\cal N} \ex^{i k_\mu x^\mu} \ , \cr
    \bar{s}(t,z) &=& s(t) + \delta {\cal S} \ex^{i k_\mu x^\mu} \ , \label{perbs}
\end{eqnarray}
where $k_\mu = (k_t,0,0,k_z)$ is the constant wave-vector for the modes and 
$\delta {\cal N}$, $\delta U^z_\n$, \etc are the constant wave-amplitudes. 
Within the setting of Eq.~(\ref{perbs}), we see that the short-wavelength 
approximation for the modes is expressible as
$k_t, k_z\gg H_i \equiv \dot A_i/A_i$ for all $i\in \{x,y,z\}$. 

The results in Figure \ref{fig:uC}, which displays the time evolution of the two 
flows $\displaystyle u_{\n,\s} = A_z u^z_{\n,\s}/u^t_{\n,\s}$ as well as the couplings 
$(\Cns,\Csn)$, show that the relevant coefficients are driven to zero by 
cosmological expansion, meaning, as expected and anticipated, that any 
instability will only operate during a finite time. Note that long before and after the 
transition, one fluid dominates over the other, so spacetime is effectively FLRW; 
only during the finite Bianchi I phase do the two fluids have comparable 
contributions \cite{cpaprd}.

\begin{figure*}[ht]
\centering
\includegraphics[height=7.0cm,width=8.0cm,clip]{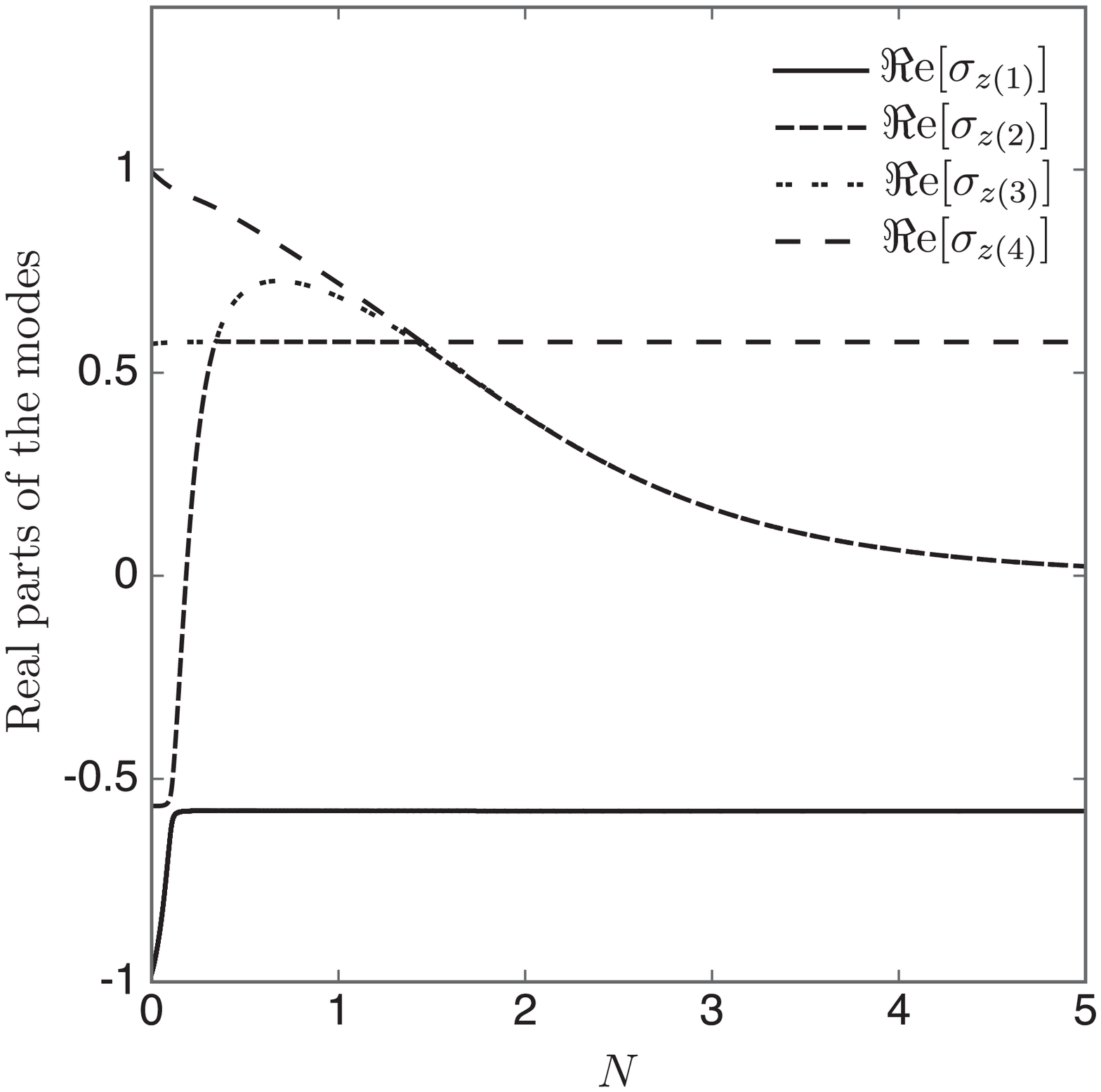}\hskip7mm
\includegraphics[height=7.0cm,width=8.0cm,clip]{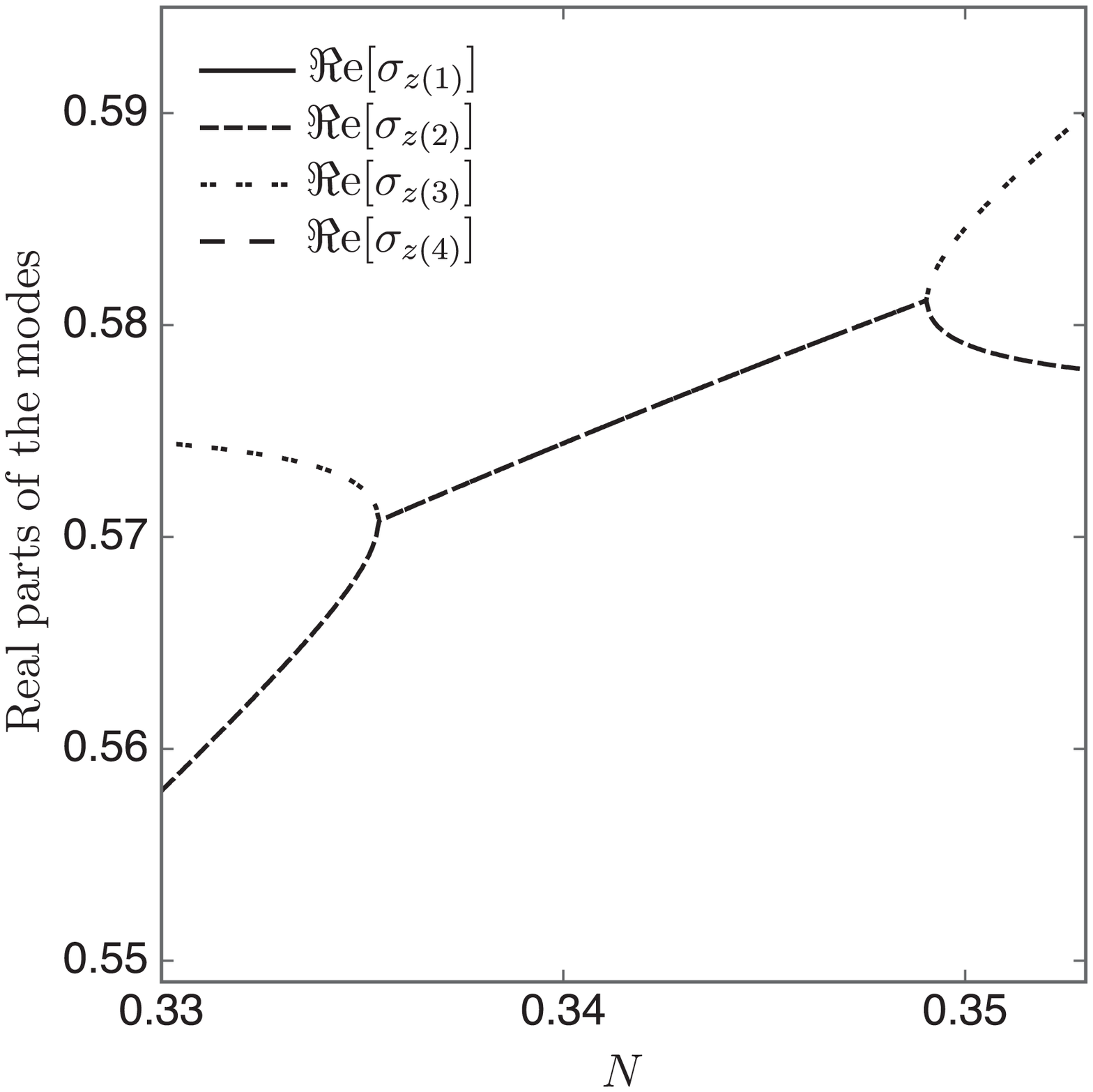}
\caption{Real parts of the perturbation mode speed $\sigma_z$ solutions of 
Eq.~(\ref{nondynccc}), as a function of $N$ throughout the transition (upper) for 
the same parameter values as in Fig.~\ref{fig:uC}. The lower figure is a 
magnification of the region where two of the modes merge (to be distinguished 
from the points where lines simply cross, since they have zero imaginary parts, 
\cf figure \ref{fig:im}).  In the limit of large $N$, one recovers asymptotically the 
four modes of constant speeds $\pm 1/\sqrt{3}$ and $0$, as expected for
radiation and matter with zero relative velocity.  The two solutions that merge 
(the lower panel) pick up imaginary components (\cf  figure \ref{fig:im}) and 
thus correspond to the epoch in which the instability initiates, develops, and 
ends.}
\label{fig:real}
\end{figure*}

Given a Bianchi I background, we place the terms of Eq.~(\ref{perbs}) into the 
Einstein and two-fluid equations of \citep{cpaprd}, expand, and keep only terms 
linear in the perturbed quantities, to arrive at the dispersion relation 
\begin{eqnarray}
   && \left[\left(u_\n \sigma_z - 1\right)^2 c^2_\n - \left(\sigma_z - u_\n\right)^2\right] 
         \left[\left(u_\s \sigma_z - 1\right)^2 c^2_\s - \left(\sigma_z - u_\s\right)^2\right] 
         \cr
   &&- \Cns \Csn \left(u_\n \sigma_z - 1\right)^2 \left(u_\s \sigma_z - 1\right)^2 = 0 
    \label{nondynccc}
\end{eqnarray}
for the mode speed $\sigma_z = - A_z k_t/k_z$.  This relation is of the exact same 
mathematical form as the dispersion relation obtained by Samuelsson \etal 
\citep{SLAC10} [\cf their Eq.~(69)].  It is thus immediately clear that our simplified 
model possesses all the ingredients for two-stream instability. 

\begin{figure}[t]
\centering
\includegraphics[height=7.0cm,width=8.0cm,clip]{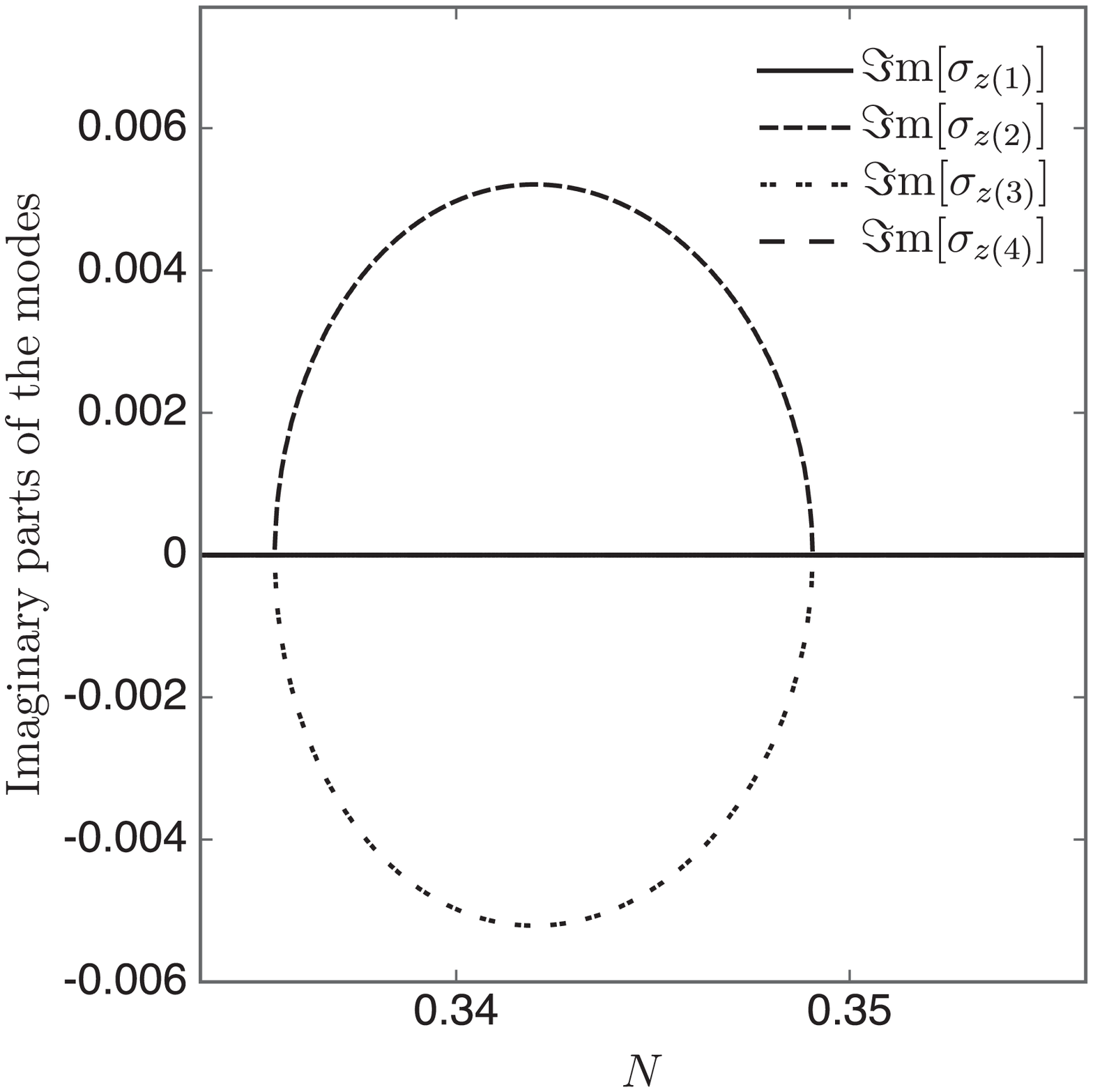}
\caption{Imaginary parts of the perturbation mode speed $\sigma_z$ solutions 
of Eq.~(\ref{nondynccc}), for the same parameter values as in Fig.~\ref{fig:real},
as a function of $N$ throughout the epoch where the modes are unstable 
(ignoring the rest of the time evolution since all the modes are real). This 
corresponds to the region in Fig.~\ref{fig:real} (lower panel) where the real parts 
become degenerate: the number of parameters needed to describe the 
perturbations is therefore constant during the transition.}
\label{fig:im}
\end{figure}

The presence of the instability is shown in Figures \ref{fig:real} and \ref{fig:im}.  
These figures provide graphs of the real and imaginary components of 
$\sigma_z$, versus the ``e-folding'' factor $N$ defined as
\begin{equation}
    N = \ln \left(A_x A_y A_z\right)^{1/3} \ .
\end{equation}
Figure \ref{fig:real} clearly shows four modes -- denoted $\sigma_{z (i)}$ with 
$i=1,\dots,4$ -- all less than one, as should be the case. These modes evolve 
asymptotically towards constants corresponding to the ``sound'' speeds expected 
for both fluids. The overall asymmetry of the modes is due to the background 
relative flow. In the lower left corner of the figure, it is interesting to note the 
presence of a so-called avoided crossing, at which two of the modes ``exchange'' 
identity.
 
The instability window is best seen in Figure \ref{fig:im} where the imaginary parts 
of $\sigma_z$ are graphed.  Consistent with this is the ``mode-merger'' of the real 
parts that occurs in Figure \ref{fig:real}. The results show that the  window opens 
and closes during the same epoch as the transition takes place. Of course, the 
window closes because of the behavior shown in Figure \ref{fig:uC}; \eg the 
relative velocity is quenched by the cosmological expansion.

It should be clear that the mechanism discussed here is robust and should apply 
equally well, for example, to a system of coupled condensates. As stressed at the 
beginning, any system with two or more fluids, and a (non-gravitational) coupling 
between them, could be subject to two-stream instability.  What is perhaps unique 
in the cosmological context is that the instability shuts down automatically, without 
any fine-tuning, so long as there is overall expansion.  

The cosmological expansion thus provides a means of both initiating and ending 
the instability. This is mostly due to the cosmological principle, which states that 
the Universe should be, at almost all times, well described by a FLRW model. 
This implies that only finite epochs can exhibit a different behavior, in the case at 
hand that of a non-isotropic Bianchi phase. There remains to address the 
question of the origin of the relative flow. One might argue, for instance, that the 
inflation era, almost by construction, drives any primordial anisotropy to zero. 
During many field inflation itself, instabilities of this sort could be spontaneously
initiated, even for a limited amount of time, and lead to new constraints.

In the companion paper \citep{cpaprd}, we suggested several scenarios that lead 
to a non-vanishing relative flow, some  being also related to the question of the 
cosmologically coherent magnetic fields which are thought to exist \citep{AK10}. 
These may actually reverse the question: it is conceivable that any model aimed 
at producing primordial magnetic fields on sufficiently large scales will also 
induce coherent anisotropies on these scales (the metric we used here is thus 
meant to describe only a cosmologically small region of space). The mechanism 
producing these magnetic fields should thus be tested against the instability 
suggested here. This question is quite natural given that two-stream instability is 
well-established for plasmas, and there is no reason why it should matter that 
these are placed in a cosmological setting.

Based on the results presented here, we suggest that cosmological two-stream 
instability should be taken into account in further studies of all the transitions that
could lead to its occurrence. In particular, the supposedly latest such transition, 
that ended the matter-domination era to the ongoing accelerated phase, could 
lead to drastically different observational predictions if the latter was driven by a 
mere cosmological constant or by a cosmic fluid \citep{ARFKA11}: this new fluid 
would have no particular reason to be aligned along the matter flow, and hence 
an instability could develop,  producing a characteristic anisotropy whose 
features still have to be investigated, at the typical scale corresponding to the 
transition. Some have  argued that such an anisotropy has already been 
measured or that it could be using forthcoming Planck data \citep{CHSS10}.

The instability demonstrated in this letter offers  a new avenue for understanding 
cosmological data, in the sense of new constraints, as well as a potential 
mechanism for generating anisotropies at specific scales and increase the tensor 
mode contribution and non-gaussianities. In this regard, much work obviously 
remains to be done before two-stream instability might be viewed as a viable, 
cosmological mechanism: we need to consider flows at arbitrary angles, include 
dissipation, how relative flows may develop, back-reaction of instabilities on the 
whole system, and so on.

Finally, we note that two-stream instability is just one example of how multi-fluid 
dynamics differs from that of a single fluid.  We have only considered a simple 
two-fluid model, with many features left out, but it still illustrates well the 
possibilities.  Perhaps the key point is that the two-stream mechanism cannot 
operate in the various one-fluid systems that are sometimes called ``multi-fluids''. 
Although these models have several different constituents, they do not account 
for relative flows. The example considered here clearly demonstrates why relative 
flows may have interesting consequences, and motivates further studies of the 
implications. The Cosmological Principle demands a frame in which all 
constituents are at rest, but we believe strict adherence is too severe, may limit 
progress, and prevent new insights into the structure and evolution of the 
Universe.   

GLC acknowledges support from NSF via grant number PHYS-0855558.  PP 
would like to thank support from the Perimeter Institute in which this work has 
been done. NA acknowledges support from STFC in the UK.

\bibliography{Bib2Streams}

\end{document}